# The citation impact outside references -
# formal versus informal citations

*Werner Marx and Manuel Cardona*
*Max Planck Institute for Solid State Research, Stuttgart (Germany)*

## Abstract

In this study the amount of "informal" citations (i.e. those mentioning only author names or their initials instead of the complete references) in comparison to the "formal" (full reference based) citations is analyzed using some pioneers of chemistry and physics as examples. The data reveal that the formal citations often measure only a small fraction of the overall impact of seminal publications. Furthermore, informal citations are mainly given instead of (and not in addition to) formal citations. As a major consequence, the overall impact of pioneering articles and researchers cannot be entirely determined by merely counting the full reference based citations.

## Introduction

The number of citations is often taken as a measure of the resonance or impact an article, a researcher or a research institute has generated up to a given date. Although the citation based impact cannot be taken as the final importance and quality of articles, citation data give evidence for strengths and shortcomings and therefore are frequently used for research evaluation involving individuals, institutions or even countries. Being cited means that a given publication (mostly a journal article, but sometimes also a publication from another document type such as a book or conference proceedings) appears as footnote or reference in the publication of another author, who refers to it for additional reading. The number of citations is seen to be a rough measure of the importance of the specific publication within the scientific community.

Citing is afflicted to a greater or lesser extent by many distortions, which have been widely discussed in the bibliometric literature [1]. However, one specific bias leading in some cases to a substantial loss of the citation based impact has not been quantified until present: the amount of "informal" citations (i.e. those mentioning only names or their initials instead of the complete references) in comparison to the "formal" (full reference based) citations. In the present study some prominent pioneers in physics and chemistry, and common name based items out of their research disciplines, are analyzed, in particular concerning the relation of their formal to informal citations.

## Methodology

The data presented here are predominantly based on the Science Citation Index (SCI) under the Web of Science (WoS), the search platform provided by the producer: Thomson Scientific (the former Institute for Scientific Information, ISI). In addition, the database SCISEARCH (SCI under STN), the literature file (CAPLUS) of the Chemical Abstracts Service (CAS) of the American Chemical Society and the Physics Abstracts file (INSPEC) have been consulted under the host STN International. The CAS file covers both chemistry and physics (physics, however, being completely covered only until around 1960) whereas the INSPEC file covers mainly physics, with some closely connected disciplines like electronics and computer science included. The SCI under the WoS, as well as the INSPEC file and the CAS literature file under STN International, stretch back to 1900. The SCI under STN covers only the time period from 1974 until present.

The Web of Science offers two basic search modes: The General Search mode and the Cited Reference Search mode. The General Search mode reveals publications in the SCI "source journals" (no books, no popular publications, no conference proceedings unless they appear in source journals). The about 6000 SCI "source journals" selected by the staff of Thomson/ISI as contributing to the progress of science represent only about 10 percent of the total world wide scientific literature, but are the dominating part [2]. The informal citations determined in this study are based both on the SCI searched under the WoS, on the CAS literature file CAPLUS and on the INSPEC file, both searched under STN International. All articles mentioning a specific name in the title, the abstract or as a keyword were considered and selected according to their year of publication. In contrast to the CAS and INSPEC file, the SCI (under WoS as well as STN) does not include the abstracts of articles published before 1991, limiting the time period of SCI informal citations which include abstracts to the time period 1991 till present.

The Cited Reference Search mode enables access to all references appeared in SCI source journals, whether the references are to articles in source journals (cited either correctly or incorrectly), or references to books or any other published material (e.g. theses, internal reports, or even private communications). By this method, the time-dependent number of citing articles (source items only) related to works of a specific author can be determined. Due to the WoS system limitations under the Cited Reference Search mode (10 pages with 500 items at maximum), the number of citing papers of some prominent pioneers (e.g. Albert Einstein) could only be determined till 1973 and was completed with respect to the time period 1974 until present by searching the SCI under STN International (using the reference author search field).

## Formal vs informal citations: scientists

The fundamental ideas of science soon find their way into textbooks. Then they are often taken for granted and scientists usually do no longer refer to the original articles. They only mention the names or the name-based items instead of the full formal references. This can be demonstrated with the impact data of Albert Einstein. The contributions of Einstein to physics are well known. His works on quantum physics and relativity are outstanding and unique. Einstein formulated the special and general theories of relativity and, in addition, he made significant contributions to quantum theory, statistical mechanics, and solid state physics. He is widely regarded as one of the greatest physicists ever. While best known for the Theory of Relativity, he was awarded the 1921 Nobel Prize for Physics for "his discovery of the law of the photoelectric effect".

The time dependent number of formal citations (references to the works of Albert Einstein) versus informal citations ("Einstein", "relativity", "relativistic" appearing in the titles, the abstracts or the keywords) was established by the procedure mentioned above. The graphs of this study express the formal citations as number of citing papers per year - instead of citations per year (see Figure 1). Please note that one citing paper may comprise more than one citation (e.g. one citing paper refers to more than one Einstein article). The citing papers compare well to the informal citations, because in both cases citations are counted only once per paper.

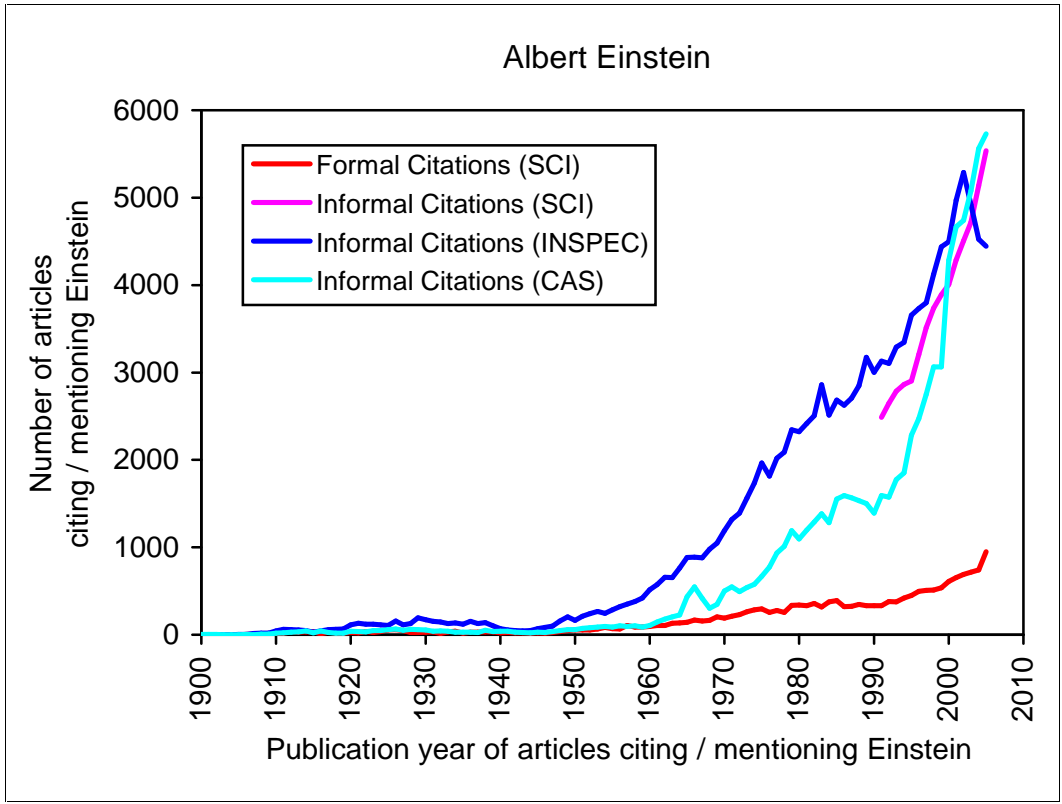

**Figure 1:** Time dependent number of formal citations related to the works of Albert Einstein versus informal citations ("Einstein", "relativity" or "relativistic" appearing in the titles, the abstracts or the keywords). Source: CAS literature file and INSPEC file under STN International and Thomson/ISI Web of Science (WoS).

Concerning Einstein, we found altogether more than 18,000 citing papers comprising about 30,000 formal citations within the time period 1900 until the end of 2005. During the same time there are almost 80,000 informal citations in the CAS file and more than 120,000 informal citations in the INSPEC file related to Einstein, based on the search terms "Einstein", "relativity" or "relativistic" appearing in the titles, abstracts or keywords (see Tables 1 and 2). Almost 30,000 informal citations are based on the titles only. From 1960 till 2005 the informal citations have outrun the formal citations by a factor of almost six (see Table 1). The CAS literature file covers more journals than SCI (see larger number of articles since the year 2000) but does not cover the complete physics literature after about 1960 (see the lower number of informal citations 1990-2000). We have collected the number of informal citations of Einstein and of the other examples so obtained and discussed in this study in Tables 1 and 2.

Some authors include full references in the abstract text. The amount of this practice (not accepted by some journal editors) was checked here with the famous article by Einstein, Podolsky, and Rosen (EPR) published 1935 (the most often cited paper of Einstein - almost 3000 citations until present). Only 4 papers mentioned the EPR paper in the abstract (see the following WoS example). All these articles cited, in addition, the actual article as formal reference.

*Title: Nonlocality of the original Einstein-Podolsky-Rosen state*
*Author(s): Cohen O*
*Source: PHYSICAL REVIEW A 56 (5): 3484-3492 NOV 1997*
*Cited References: 31      Times Cited: 12*
*Abstract: We examine the properties and behavior of the original Einstein-Podolsky-Rosen (EPR) wave function [Phys. Rev. 47, 777 (1935)] and related Gaussian-correlated wave functions. We assess the degree of entanglement of these wave functions and consider …*

Google and Google Scholar have become additional powerful WEB-search resources that we have not discussed so far. Searching the WEB for "Einstein" with Google results in the overwhelming number of 67 million entries, while Google Scholar results in 0.5 million hits (see Table 3). The Google searches were carried out without any limitations concerning format, language, and time.

We have analyzed some other pioneers of quantum physics with respect to formal versus informal citations: Erwin Schrödinger achieved fame for his contributions to quantum mechanics, especially the Schrödinger equation, for which he received the Nobel Prize in 1933. Paul Adrien Maurice Dirac pioneered the theory of the relativistic electron. He formulated the so-called Dirac equation, which led to the prediction of the existence of antimatter. Dirac shared the Nobel Prize in physics for 1933 with Erwin Schrödinger.

Schrödinger's informal citations result mainly from mentioning the Schrödinger equation, which has become what is probably the most fundamental mathematical contribution to quantum physics. His formal citations are only about 20 percent of the informal ones, which outrun the formal citations since 1960 (see Figure 2). In contrast to Schrödinger, both citation versions related to Dirac are equal until around 1980 (see Figure 3). In the last three decades, however, Dirac's informal citations have exceeded the formal citations by almost a factor of two.

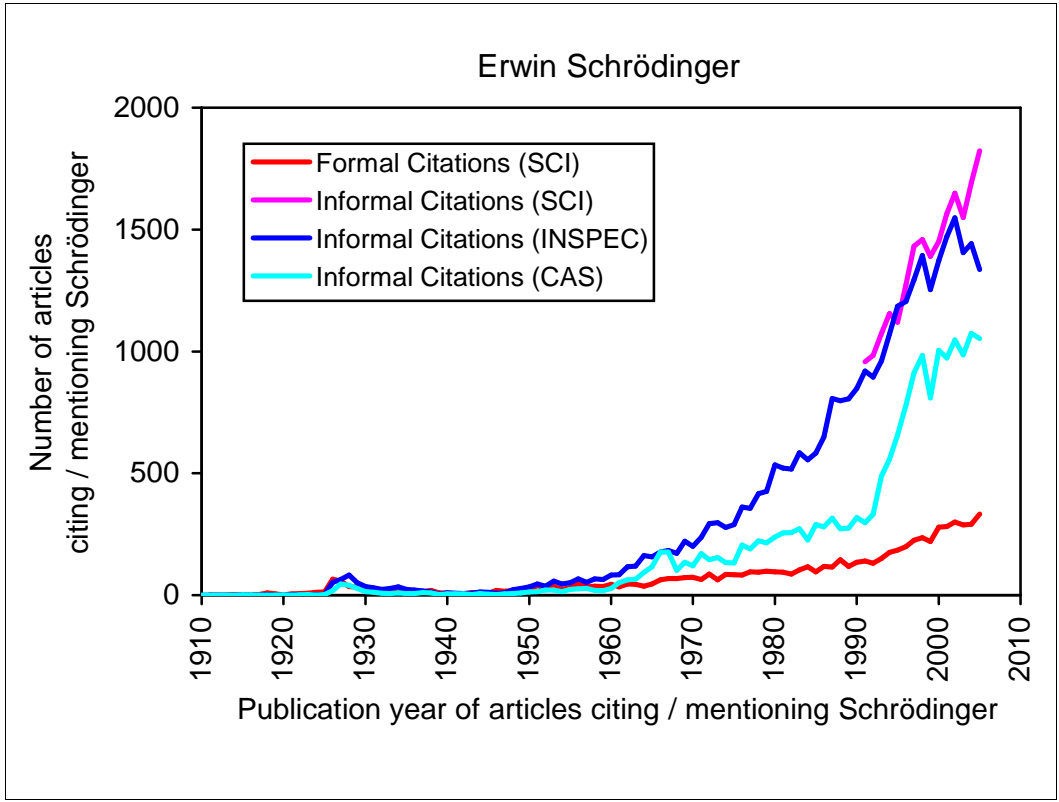

**Figure 2:** Time dependent number of formal citations related to the works of Erwin Schrödinger versus informal citations ("Schroedinger" or "Schrodinger" appearing in the titles, the abstracts or the keywords).

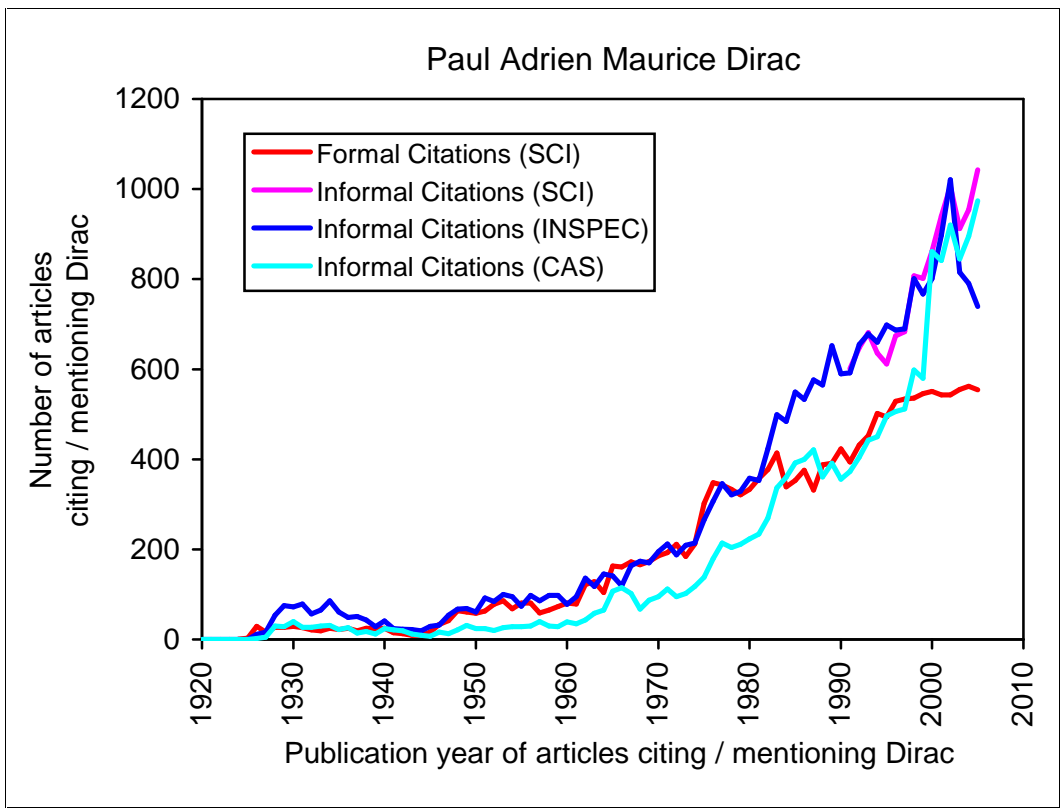

**Figure 3:** Time dependent number of formal citations related to the works of P.A.M. Dirac versus informal citations ("Dirac" appearing in the titles, the abstracts or the keywords).

Another early pioneer is Sir Chandrasekhara Venkata Raman, an Indian physicist who was awarded the 1930 Nobel Prize in Physics for his experimental discovery of the inelastic scattering of light, now generally known as Raman effect. The widely used Raman spectroscopy is based on this phenomenon. Widespread use of this experimental technique after the introduction of the laser in 1962 resulted in a strongly increasing number of informal citations with more than 8000 papers in the year 2005 (see Figure 4). The number of formal citations, however, remained constant at a level of about 50 citations per year for decades, which amounts to only about half a percent of the articles mentioning Raman's name (see Tables 1 and 2).

Until the year 2005 the overall number of formal citations to Raman's work is about 2200, whereas informal citations have reached almost 150,000 based on the CAS file and about 90,000 based on the INSPEC file. More than 50,000 articles with Raman's name appearing in the titles were published in the journals covered by the SCI. The overwhelming number of informal citations in combination with the large ratio of informal to formal citations indicates that Raman soon became a household word after the discovery of his effect.

Most of these citations (formal as well as informal) are related to the Raman effect (Raman spectroscopy, Raman scattering) [3,4]. Raman spectroscopy has become a standard powerful experimental technique with widely available commercial instrumentation. It finds applications in physics, chemistry, mineralogy, engineering, biology, environmental sciences, pharmacy, and medicine. At the same time it is a very fruitful field of basic research, both theoretical and experimental. A biannual series of an international and interdisciplinary conference is devoted to this field (the last one, ICORS 20, has just been held in August 2006 in Yokohama, with a participation of about 650 scientists).

At this point one should mention, however, that the effect was independently discovered almost simultaneously with Raman by Landsberg and Mandelstam in Moscow [5] (the priority is disputed; in the Russian literature Raman scattering is referred to as "combination scattering" - according to WoS combination scattering has been cited informally 324 times). In spite of Raman's priority concerning the experimental discovery of the effect, the Austrian scientist A. Smekal predicted it on theoretical grounds as far back as 1923 [6]. Hence, sometimes one finds in the literature informal citations to the Smekal-Raman effect. The Smekal article [6] has been cited formally about 100 times. The fact that most citations to Smekal's prediction are formal indicates that "Smekal" did not become a household word.

Because of the widespread use of Raman scattering we searched the WEB using Google: We obtained almost 14 million entries for "Raman" and 1,640,000 entries for "Raman Scattering" (see Table 3). Under "Smekal Raman" we found only about 500 Google hits.

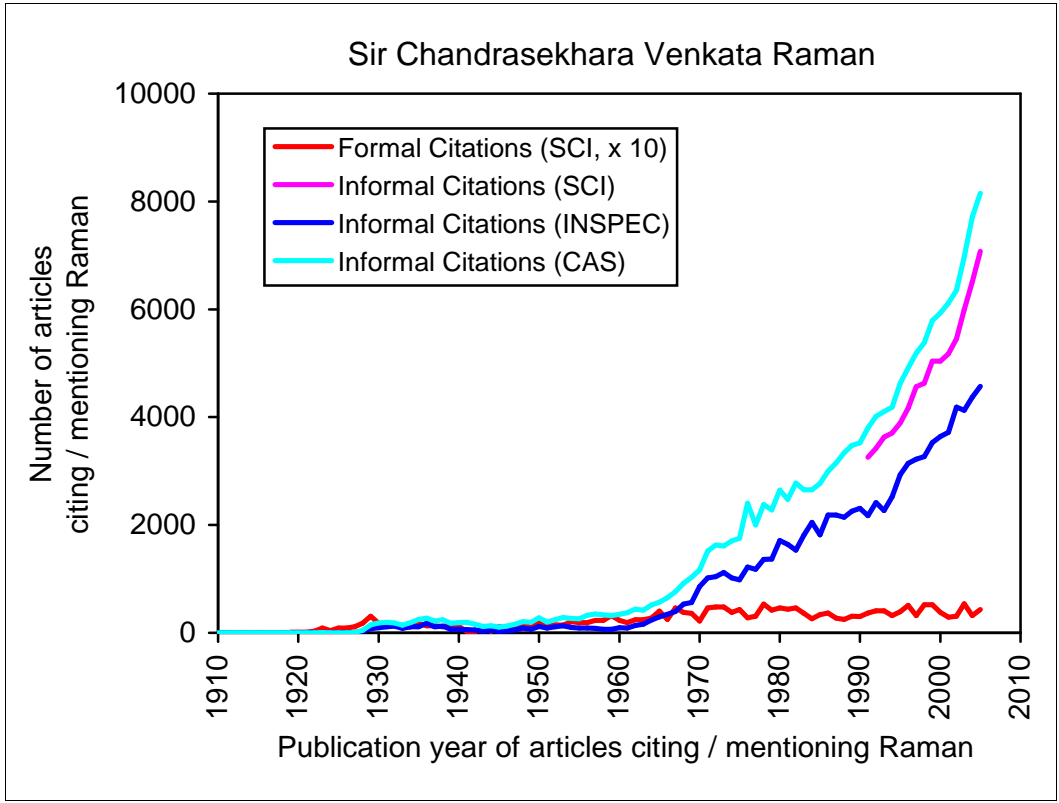

**Figure 4:** Time dependent number of formal citations related to the works of Sir C.V. Raman versus informal citations ("Raman" appearing in the titles, the abstracts or the keywords).

Another widely used spectroscopic technique, closely related to Raman spectroscopy, is Brillouin spectroscopy. It was predicted by Leon Brillouin just before the beginning of WW I [7]. Brillouin has been cited informally almost 19,000 times as opposed to about 6000 formal citations within the time period 1900 until the end of 2005. Of the informal citations 18,700 simply refer to Brillouin, about 7000 to Brillouin zone, a most important concept when dealing with spatially periodic systems, about 6500 refer to Brillouin scattering and almost 500 to Brillouin spectroscopy. Most informal citations refer to the Brillouin effect (plus scattering and spectroscopy), however, about 5400 informal citations refer to the WKB method to solve the Schrödinger equation, to be discussed later.

Although Brillouin scattering was predicted by Brillouin in 1914 [7], its discovery (by E. Gross in Leningrad) had to wait till 1930 [8]. At that time the observation of the Brillouin effect with extant incoherent light sources was a real "tour de force" and Brillouin spectroscopy had to wait till the advent of the laser in order to reach its full capabilities. The laser, together with multipass Fabry Perot spectrometers, brought Brillouin spectroscopy to its full capabilities around 1970, making the name of Brillouin also a household word. Since about 1965 the informal citations to Brillouin outrun the formal citations, exceeding the formal citations by a factor of about four (see Figure 5). The Brillouin article has been cited formally almost 200 times.

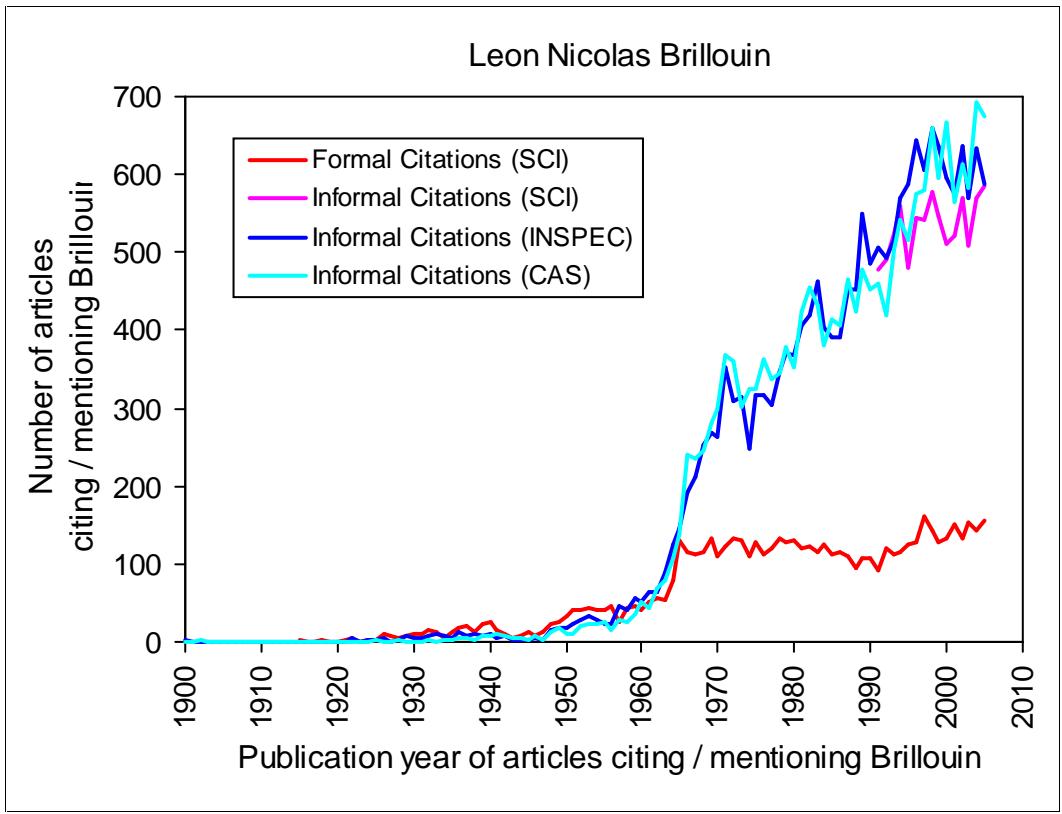

**Figure 5:** Time dependent number of formal citations related to the works of L.N. Brillouin versus informal citations ("Brillouin" appearing in the titles, the abstracts or the keywords).

## Formal vs informal citations: effects, theories, and acronyms

Informal citations sometimes comprise names in combination with terms like "approximation", "equation", "law", "theory", "spectroscopy" etc. as complex items. Such informal citations named after two or more researchers usually appear as acronyms. A typical example is the BCS theory (named for its creators: Bardeen, Cooper, and Schrieffer) [9,10]. This work, a breakthrough in theoretical solid state physics, explains conventional superconductivity, i.e. the ability of certain metals to conduct electricity at low temperatures without electrical resistance. The key idea is that electron phonon coupling results in an attractive interaction leading to bound electron pairs (so-called Cooper pair(s), which also appear as informal citations).

Whereas the 1957 article by Bardeen, Cooper and Schrieffer [9] contains the full mathematical development of the BCS theory, the formation of Cooper pairs had already been predicted a year earlier in an article with L.N. Cooper as sole author [10]. This article has been formally cited almost 500 times whereas the total number of informal citations of "Cooper pair(s)" according to INSPEC is 4000. If we confine ourselves to the title words of articles covered by the SCI we find "Cooper pair(s)" informally cited about 400 times.

The formal citations of the BCS paper published 1957 (almost 5000 citations till present) peaked some years after the publication and then decreased slowly (see Figure 6). Since the discovery of the high-temperature superconductivity in the year 1986 by Bednorz and Mueller [11], the BCS paper has been formally cited at a nearly constant rate of about 120 citations per year (except for a weak jump around 1987). The informal citations, however, increased strongly since the BCS discovery, exceeding altogether more than 10,000 until the year 2005 (see Table 1). Only a minor fraction of them contains the original BCS paper as a full reference.

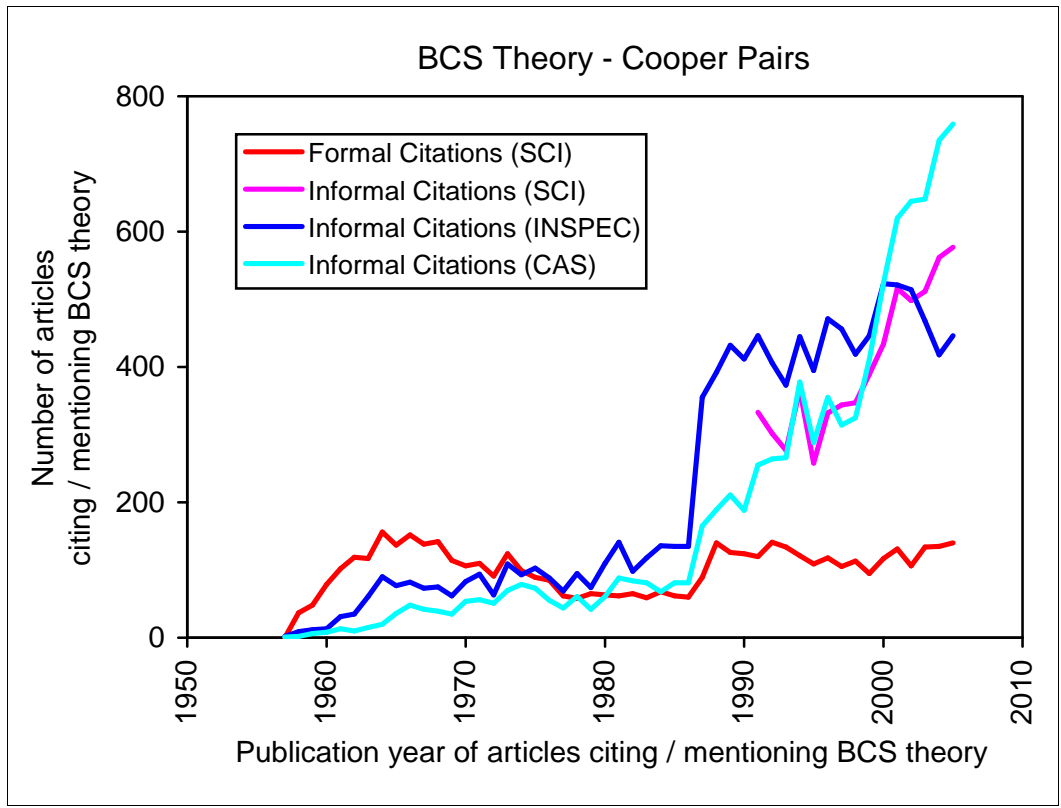

**Figure 6:** Time dependent number of formal citations related to the original BCS paper versus informal citations ("Bardeen Cooper Schrieffer" or "BCS" or "Cooper pair/s" appearing in the titles, the abstracts or the keywords).

A typical example for a method or theory not named after its creators is the Density Functional Theory (DFT). This is a quantum mechanical algorithm used in physics and chemistry to simplify the investigation of the electronic structure of many-body systems, in particular molecules and condensed phases. DFT is among the most popular and versatile methods available in condensed matter physics and computational chemistry. For his development of the "density-functional theory" Walter Kohn was awarded the 1998 Nobel Prize in Chemistry.

The original DFT paper by Kohn and Sham [12] has been cited almost 12,000 times until present and is thereby one of the most-cited articles in either chemistry or physics. In contrast to the usual citation time pattern, this paper has not yet reached its peak but has continued to increase its impact monotonically with time. Nevertheless, since the year 1993 the informal citations outrun the formal citations exceeding about 7000 citations in the year 2005 (see Figure 7). The overall number of informal citations in the CAS file within the time period 1965 till 2005 is around 40,000 (see Table 1).

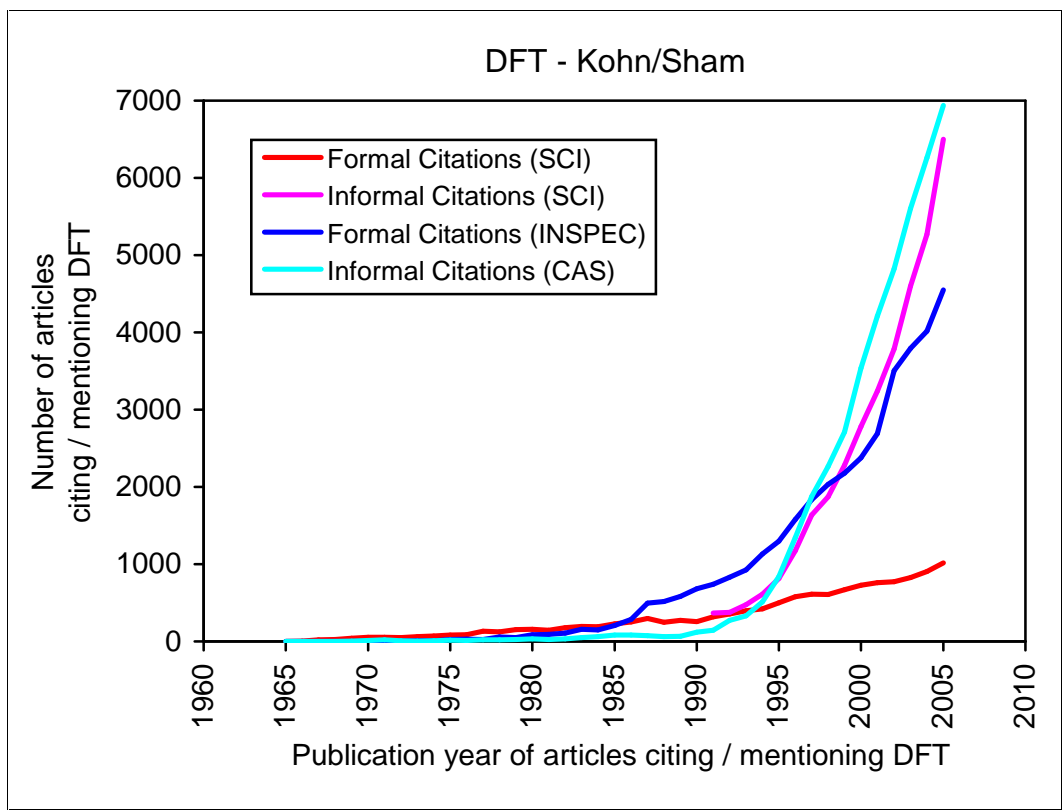

**Figure 7:** Time dependent number of formal citations related to the original DFT article versus informal citations ("Density Functional Theory" or "DFT" appearing in the titles, the abstracts or the keywords).

The Quantum Hall Effect (often referred simply as the QHE) is a quantum-mechanical version of the Hall effect [13], observed in two-dimensional electron systems at low temperatures and under strong magnetic fields, in which the Hall conductance σ takes on discrete values. The QHE was discovered by K. von Klitzing, G. Dorda, and M. Pepper in the year 1980 - Klaus von Klitzing was awarded the 1985 Nobel Prize in Physics for his experimental discovery and theoretical interpretation [14]. The citation impact of the original paper by von Klitzing peaked around 1986 and then reached a constant level of 60 formal citations until present. The QHE informal citations in the CAS and the SCI file, however, increased until the year 1998 then leveling off to the large number of about 400 citations per year (see Figure 8).

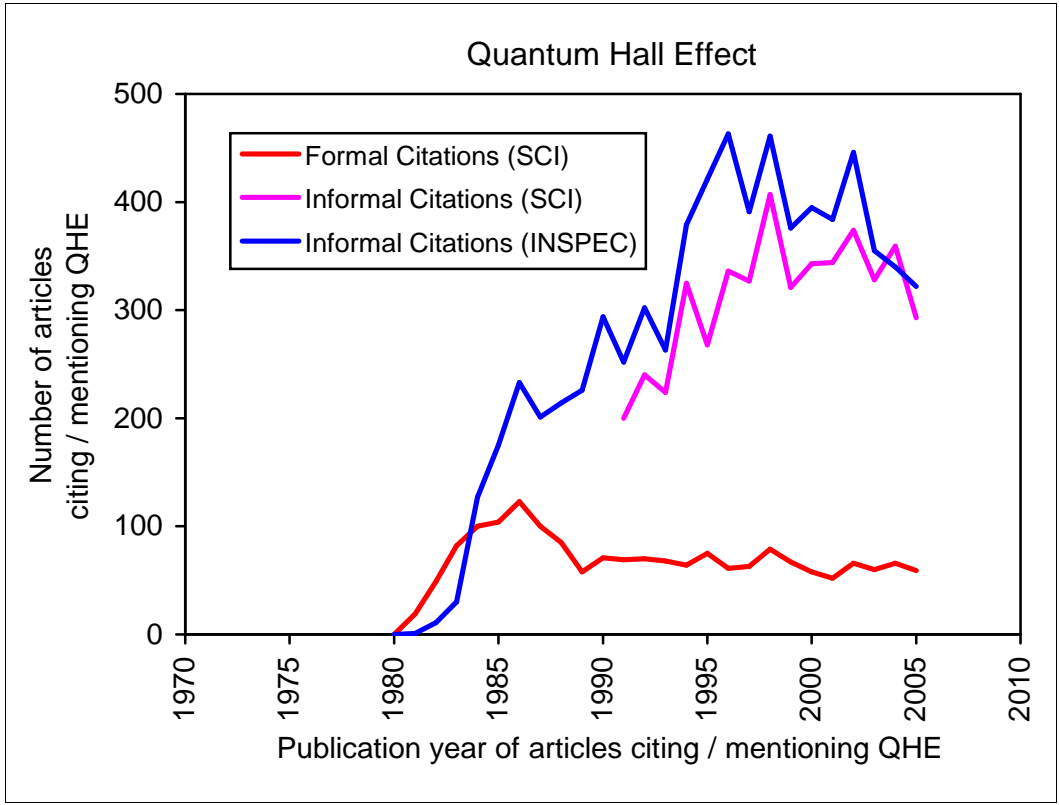

**Figure 8:** Time dependent number of formal citations related to the Quantum Hall Effect versus informal citations ("Quantum Hall Effect" or "QHE" appearing in the titles, the abstracts or the keywords).

Moore's law is the empirical observation that the complexity and/or density of integrated circuits, with respect to component cost, doubles every 18 months. It is attributed to Gordon E. Moore, a co-founder of the Intel company. Moore's original statement can be found in his publication entitled "Cramming more components onto integrated circuits" [15]. It is interesting that the formal citations still outdo the informal ones (see Figure 9). However, the law is sometimes paraphrased without mentioning it exactly as "Moore's law" (e.g. "Moore's prediction").

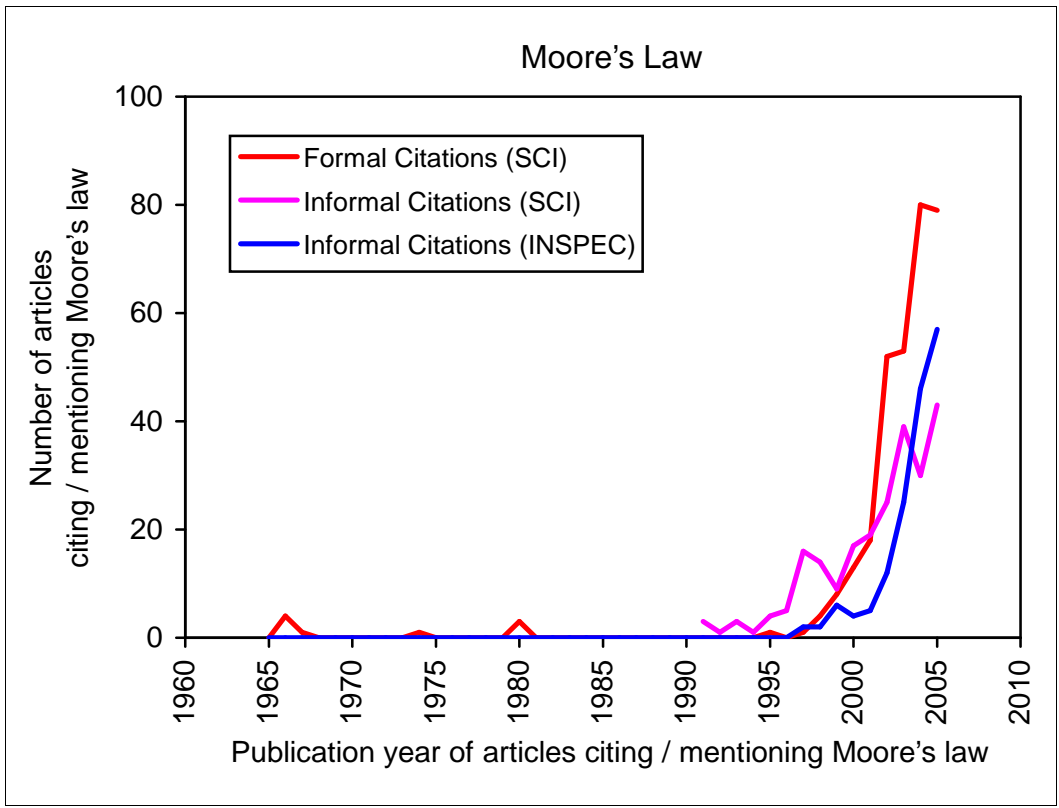

**Figure 9:** Time dependent number of formal citations related to Moore's law versus informal citations ("Moore's law" or "Moores law" appearing in the titles, the abstracts or the keywords).

Schrödinger's cat is a seemingly paradoxical thought experiment (Gedankenexperiment), devised by Erwin Schrödinger, which attempts to illustrate the incompleteness of quantum mechanics. The original article appeared in the German magazine Naturwissenschaften [16-18]. It was intended as a discussion of the EPR article published by Einstein, Podolsky, and Rosen in the same year. The number of formal citations at least equals the number of informal citations based on the SCI and the INSPEC file. (see Figure 10). Like Moore's law, Schrödinger's cat may also be paraphrased and thus not be fully covered by this search.

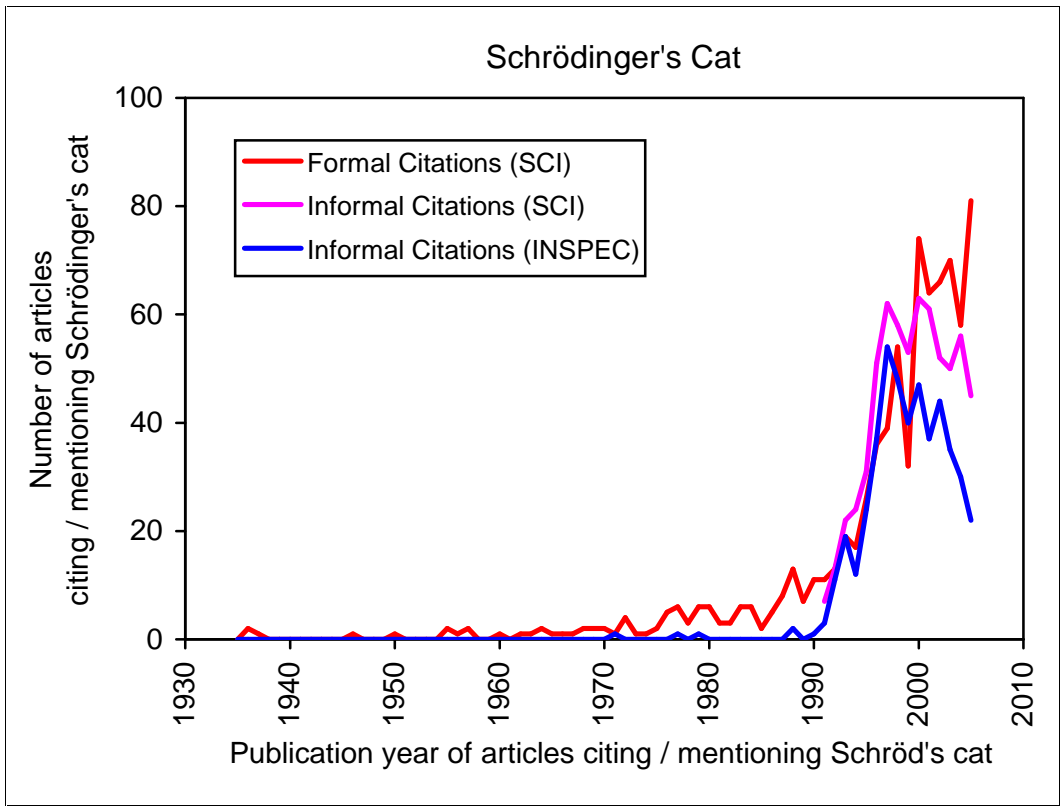

**Figure 10:** Time dependent number of formal citations related to Schrödinger's cat versus informal citations ("Schro(e)dinger's cat" or "Schro(e)dingers cat" appearing in the titles, the abstracts or the keywords).

In theoretical physics, the WKB (Wentzel-Kramers-Brillouin) approximation, also known as WKBJ approximation (J stands for Jeffreys), or a permutation of those four letters, is the most familiar example of a semi-classical calculation in quantum mechanics in which the wave function is recast as an exponential function, semi-classically expanded, and then either the amplitude or the phase is taken to be slowly varying in space. It represents an expansion of the wave function. This method is named after physicists Wentzel, Kramers, and Brillouin, who developed it in the year 1926 [19-21]. In 1924, mathematician Harold Jeffreys had already developed a general method of finding approximate solutions to linear, second-order differential equations, which include the Schrödinger equation [22]. But since the Schrödinger equation was developed two years later, and Wentzel, Kramers, and Brillouin were apparently unaware of this earlier work, Jeffreys is often not given proper credit. Articles in quantum mechanics contain any number of combinations of the three or four initials.

At the date of our WKBJ search (14.09.06) 425 articles cited at least 1 of the 4 original papers as a full reference - only 40 articles cited all 4 paper simultaneously. The number of informal citations of the WKBJ approximation is about 5400 (see Table 1). Out of these informally citing papers 80 articles cited simultaneously at least 1 of the 4 original papers - only 14 articles cited simultaneously all 4 original papers. The informal citations strongly increased since 1960 reaching at present 200 citations per year (see Figure 11).

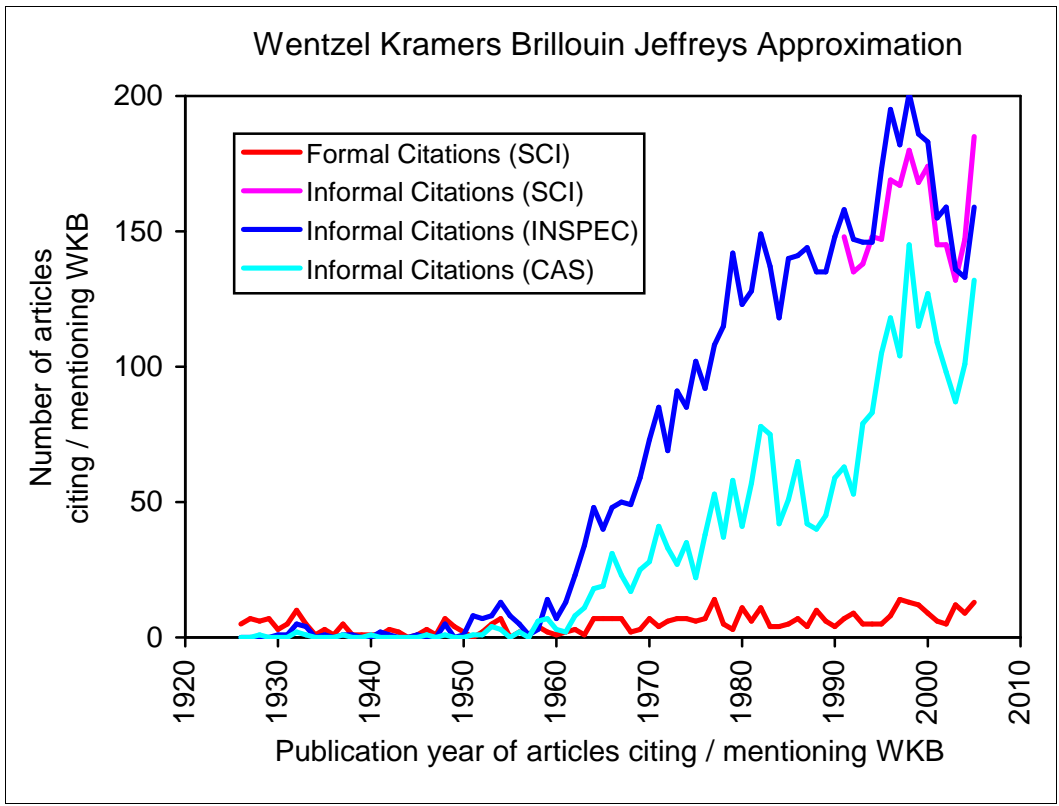

**Figure 11:** Time dependent number of formal citations related to the WKBJ approximation versus informal citations (("Brillouin" AND "Kramers" AND "Wentzel/Wenzel") OR "BKW" OR "KBW" OR "BWK" OR "KWB" OR "WBK" OR "WKB" OR "WKBJ" OR "JWKB" appearing in the titles, the abstracts or the keywords).

**Early scientists with high informal to formal citation ratios**

The Doppler effect, named after Christian Andreas Doppler, is the change in frequency and/or wavelength of a wave that is perceived by an observer moving relative to the source of the waves. For waves such as sound waves, which propagate in a material medium, the effect depends on whether the source or the observer move with respect to the propagation medium. In the case of light, according to relativity theory the effect depends on the relative velocity of observer and medium. Christian Andreas Doppler, born in Salzburg (Austria) in 1803, occupied several minor academic positions till he was appointed professor of geometry and mathematics at the Polytechnical School of Prague in 1841. Most of his publications are nearly forgotten, except for that in which he predicted the effect that bears his name [23]. He developed the theory of his effect in order to explain the color of double stars related to their rotation around each other. The color of the star that approaches the observer should be blue shifted, that of the one, which moves away, should be red shifted.

We have only found in the WoS about 100 formal citations to the report in which the Doppler effect was first published [23]. Figure 12 displays the informal citations to "Doppler" as obtained under the SCI (as of 1990 so as to include abstracts) and of INSPEC (going back to 1990 with abstracts included). Doppler was cited informally more than 80,000 times according to the SCI file and almost 47,000 times according to INSPEC. As often the case in previous examples, the number of citations took off around 1950, reaching the amount of 1200 per year in 1990 according to INSPEC and almost 5000 per year according to the SCI.

It is interesting to compare in the case of Doppler the INSPEC informal citation data with the SCI data after 1991 (see Figure 12). The latter are about 2.5 times higher than the former. The explanation of this difference is in fact mainly due to the fact that the SCI includes medical literature. Doppler became a household word among medical personnel shortly after 1973, in connection with the invention and development of Doppler sonography, a technique that allows the noninvasive observation of the circulation of blood in the main blood vessels [24]. The term "Doppler sonography" first appears in the SCI as informal citation in 1973 (2 times), rising rapidly and reaching about 350 in 2005. Beside Green's function (see below) Doppler shows the largest ratio of informal to formal citations (see Tables 1 and 2).

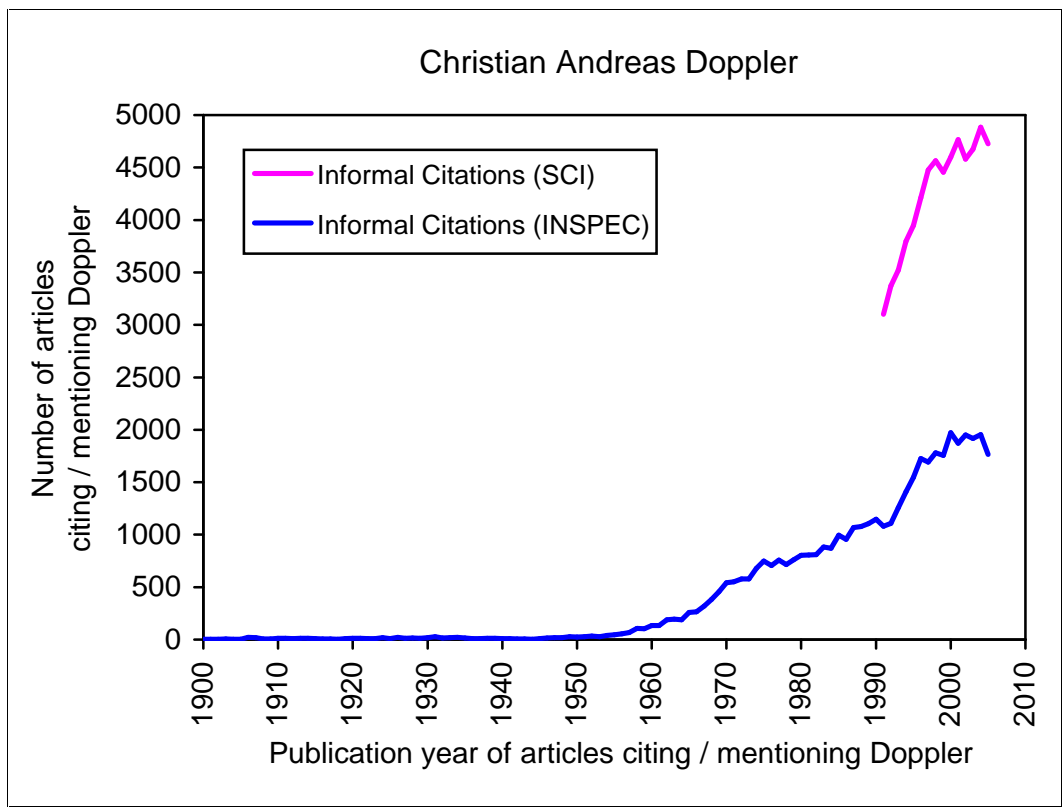

**Figure 12:** Time dependent number of formal citations related to "Doppler" appearing in the titles, the abstracts or the keywords.

Informal citations are rather widespread for publications dating back to the early days of science. Very often, the journals to be cited are not readily available but scientists know the name and subject of the work, which propagates from one article to the next without mentioning the precise (formal) source. As an example, we discuss George Green, a Miller from Nottingham, who in 1828 published an article where he introduced the Green function, sometimes mentioned as Green's function, as a tool for solving differential equations. The original work appeared in 1828 in a rather obscure publication of the Nottingham Subscription Library with the Title "An Essay on the Application of Mathematical Analysis to the Theories of Electricity and Magnetism" [25].

Such complicated titles are prone to formal citation errors. Indeed we find a total of only 30 formal citations to Green's original paper with the source involving several permutations of the title above. When looking at the informal citations (see Figure 13) one must take into account that they appear under either Green function or Green's function (the more frequent version) and that one must accumulate the singular and the plural of the word "function". The number of informal citations is overwhelming (see Tables 1 and 2), especially considering the fact that they reflect a single publication which has been formally cited only 29 times (probably because of the unavailability of such an obscure publication).

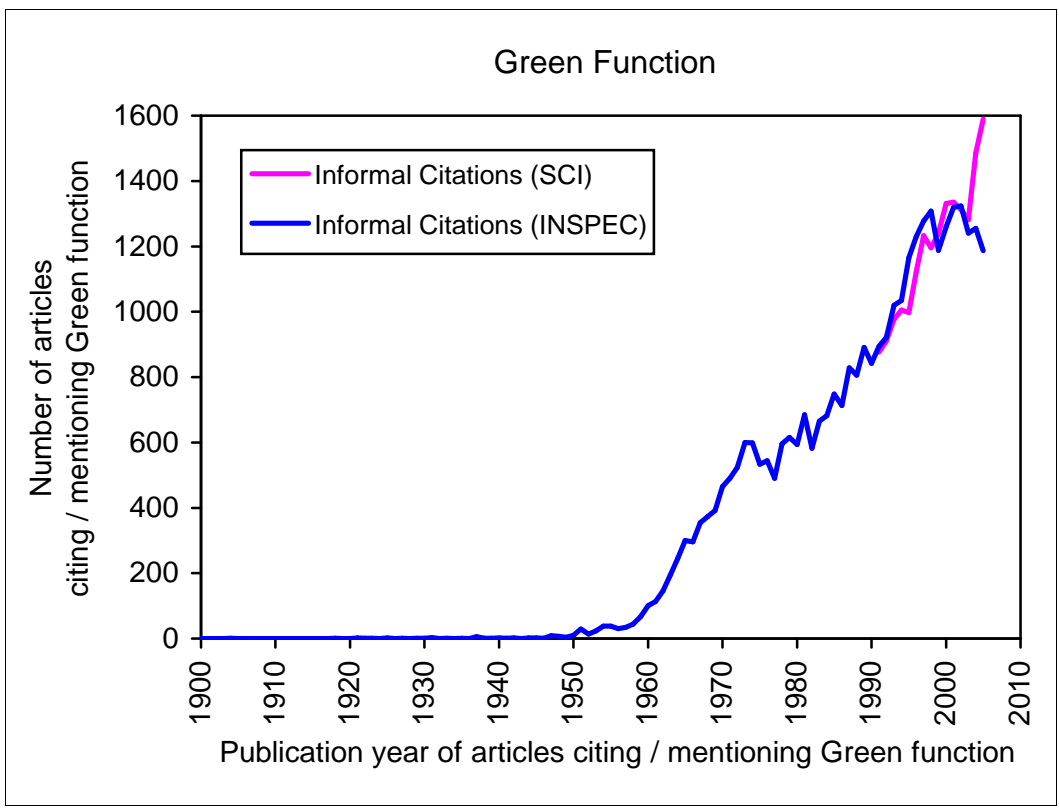

**Figure 13:** Time dependent number of informal citations related to the Green function ("Green(s) function(s)" or "Green's function(s)" appearing in the titles, the abstracts or the keywords).

### Less known scientists

Harry Dember, born in 1882 in Mansfeld (Thuringia) is best known for the discovery of the "Dember effect" as reported in three articles in the Physikalische Zeitschrift [26-28]. The first of these articles [26] describes the observation of the effect on natural cuprite ($Cu_2O$), at that time a canonical semiconducting material. The effect is seen as a voltage, which appears along the direction of light incident on one of the sample faces. Dember interpreted the effect as due to the electrons excited by the absorbed light (it has been realized later that the Dember effect involved the diffusion of electrons as well as holes). The other two articles [27-28] report experiments performed to demonstrate that the effect is a volume effect and not an effect resulting from charge separation at blocking layers or other inhomogeneities. Photovoltaic effects known at that time had been shown to be related to such layers.

In 1923, Dember was appointed professor of physics at the university of Dresden. He had till then published 13 papers in SCI source journals which to date have not received much attention (a total of 24 citations only). His major contribution is the discovery of the Dember effect appearing 1931 in the Physikalische Zeitschrift [26] (144 formal citations) and in the two additional articles appeared in 1931 and 1932 [27-28] (75 and 83 citations). His career as an author ended in 1933 with his emigration first to Istanbul, in order to escape Nazi persecution, followed by the acceptance, in 1942, of a lectureship at Rutgers University in the USA, where he passed away in 1943 [29]. His three major contributions have been cited rather regularly at the average rate of about 6 times per year until present.

Since these publications were written in German and appeared in a journal not easily available during WW II and even today, it is interesting to investigate the informal citations this work has received. At the date of our search (14.11.06) the word "Dember" appears in titles and abstracts of SCI source articles 135 times, of which 86 correspond to "Dember effect", 26 to "Dember field", 6 to "Dember model" and 21 to "photo-Dember". In the case of Dember, only 20 % of the informally citing articles give also a formal citation to one of the three papers under discussion in connection with the Dember effect [26-28]. This example shows that even in the case of low profile authors, with only a few highly cited contributions, informal citations (135 in SCI and 331 in INSPEC) contribute considerably to his impact and must be taken into account when determining either the author's impact or that of his main publications.

In most cases studied here the number of informal citations based on the SCI within the time period 1991 till 2005 (with the abstracts included) compares well with those based on the CAS literature file and the INSPEC file. The difference between the CAS and the INSPEC file results mainly from the incomplete coverage of chemistry under the INSPEC file (see DFT and Raman) and of physics (since around 1960) under the CAS file (see Einstein, Schrödinger, and WKBJ). The somewhat larger numbers based on the CAS and the INSPEC file as compared with those under the SCI file is most likely due to the fact that the SCI (under the General Search mode) only reports citations appearing in source articles within the selected SCI source journals. The decrease of informal citations since 2003 based on the INSPEC file is presumably caused by delayed input in comparison to the SCI. Note also that the long-lasting exponential growth of science has brought about a large excess of weight concerning more recent literature: The records covered both by the SCI and the CAS file increased approximately by a factor of hundred within the 20th century.

Tables 1 and 2 summarize the impact data shown above. The tables include some additional names not presented with the detailed time curves of their impact. Please note that the informal citations given in Table 1 are based on the INSPEC file instead of the SCI file as in Table 2. For comparison and completeness, we have displayed in Table 3 informal citation data searched under Google and Google Scholar.

**Table 1:** Number of formal citations (SCI) and informal citations (INSPEC) in the time period 1900-2005.

| name | # formal citations (SCI) | # informal citations (INSPEC) | informal citations / formal citations |
|---|---|---|---|
| Bose | 2440 | 12 929 | 5.3 |
| Brillouin | 6110 | 18 700 | 3.1 |
| Dirac | 17 129 | 22 813 | 1.3 |
| Einstein | 18 067 | 123 196 | 6.8 |
| Feynman | 29 881 | 10 852 | 0.4 |
| Heisenberg | 7373 | 17 675 | 2.4 |
| Planck | 2870 | 14 496 | 5.1 |
| Raman | 2210 | 146 887* | 66.5 |
| Schrödinger | 6890 | 31 724 | 4.6 |
| BCS Theory / Cooper Pairs | 4967 | 10 704 | 2.2 |
| Blonder Tinkham Klapwijk | 1204 | 108 | 0.1 |
| Dember Effect | 144 | 331 | 2.3 |
| Doppler Effect | 100 | 46 560 | 466 |
| Density Functional Theory | 12 844 | 42 509* | 3.3 |
| Feynman Diagram | 905 | 4367 | 4.8 |
| Fokker Planck Equation | 114 | 7792 | 68.4 |
| Green's Function | 29 | 34 013 | 1173 |
| Quantum Hall Effect | 1768 | 7062 | 4.0 |
| Moore's Law | 318 | 159 | 0.5 |
| Ruderman Kittel | 2682 | 637 | 0.2 |
| Schrödinger's Cat | 783 | 469 | 0.6 |
| Wentzel Kramers Brillouin | 418 | 5381 | 12.9 |

* based on the CAS literature file

A numerical comparison of formal versus informal citations is somewhat problematic because the informal citations appearing in the abstracts of the publications are not available before 1991 in the SCI. The citation data given in Table 2 correspond only to the year 2005, indicating the current informal/formal relation.

**Table 2:** Number of formal and informal citations (both SCI) appearing only during the year 2005.

| name | # formal citations (SCI) | # informal citations (SCI) | informal citations / formal citations |
|---|---|---|---|
| Bose | 250 | 1251 | 5.0 |
| Brillouin | 156 | 149 | 1.0 |
| Dirac | 554 | 1043 | 1.9 |
| Einstein | 951 | 5535 | 5.8 |
| Feynman | 1109 | 353 | 0.3 |
| Heisenberg | 207 | 885 | 4.3 |
| Planck | 96 | 714 | 7.4 |
| Raman | 430 | 7074 | 16.5 |
| Schrödinger | 333 | 1823 | 5.5 |
| BCS Theory / Cooper Pairs | 140 | 577 | 4.1 |
| Blonder Tinkham Klapwijk | 89 | 11 | 0.1 |
| Density Functional Theory | 1017 | 6498 | 6.4 |
| Dember Effect | 2 | 10 | 5.0 |
| Doppler Effect | 3 | 4726 | 1575 |
| Feynman Diagram | 17 | 85 | 5.0 |
| Fokker Planck Equation | 3 | 302 | 100 |
| Green's Function | 4 | 1589 | 397 |
| Quantum Hall Effect | 59 | 293 | 5.0 |
| Moore's Law | 79 | 43 | 0.5 |
| Ruderman Kittel | 53 | 57 | 1.1 |
| Schrödinger's Cat | 81 | 45 | 0.6 |
| Wentzel Kramers Brillouin | 13 | 185 | 14.2 |

**Table 3:** Number of informal citations under Google and Google Scholar (without any limitations concerning format, language, and time).

| name | # informal citations (Google) | # informal citations (Google Scholar) | # Google / # Google Scholar |
|---|---|---|---|
| Brillouin | 1 960 000 | 158 000 | 12 |
| Dirac | 6 350 000 | 314 000 | 20 |
| Einstein | 67 100 000 | 498 000 | 135 |
| Feynman | 6 980 000 | 163 000 | 43 |
| Heisenberg | 6 040 000 | 180 000 | 34 |
| Planck | 21 600 000 | 401 000 | 54 |
| Raman | 13 700 000 | 760 000 | 18 |
| Schrödinger | 3 420 000 | 58 200 | 59 |
| Dember | 95 900 | 2320 | 41 |
| Doppler | 23 000 000 | 773 000 | 30 |
| Dember Effect | 602 | 319 | 2 |
| Doppler Effect | 949 000 | 23 100 | 41 |
| BCS Theory | 212 000 | 10 100 | 21 |
| Bardeen Cooper Schrieffer | 54 900 | 2 790 | 20 |
| Cooper Pairs | 446 000 | 12 200 | 37 |
| Blonder Tinkham | 10 300 | 500 | 21 |
| Density Functional Theory | 1 780 000 | 99 000 | 18 |
| Feynman Diagram | 246 000 | 10 700 | 23 |
| Fokker Planck Equation | 619 000 | 26 700 | 23 |
| Quantum Hall Effect | 920 000 | 13 000 | 71 |
| Green's Function | 1 410 000 | 13 300 | 106 |

| Moore's Law | 1 420 000 | 4280 | 5 |
|---|---|---|---|
| Ruderman Kittel | 45 400 | 3 400 | 13 |
| Schrödinger's Cat | 392 000 | 662 | 592 |
| Wentzel Kramers Brillouin | 27 500 | 2 090 | 13 |
| WKB Approximation | 190 000 | 9980 | 19 |
| WKBJ Approximation | 913 | 513 | 1 |

Date of Google search: October/November 2006

**Conclusions**

The number of informal citations is often many times higher than the number of formal citations, especially when the name of an author or his contributions have become household words. This is also the case, when work appeared in old, obscure, or not easily available journals. Hence, the formal citations measure only a fraction of the (measurable) overall impact of seminal papers or pioneer authors like those presented here. Only a small fraction of the total number of the citing papers (formal plus informal citations) cite simultaneously the full reference and the author's name. Thus, informal citations are mainly carried out instead of (and not in addition to) formal citations. As a major consequence, citation rankings of pioneers may be greatly misleading. The overall impact of pioneering articles cannot be entirely determined by merely counting their citations.